# AN OVERVIEW OF VARIOUS BIOMETRIC APPROACHES: ECG AS ONE OF ITS TRAIT

**Kavyashree .U[1], K.N Deeksha[2], Suma Ballal[3], Vitina Mary D'souza[4], Rama Moorthy H[5]**

[1]Student, CSE, SMVITM-Bantakal, Karnataka, India

[2]Student, CSE, SMVITM-Bantakal, Karnataka, India

[3]Student, CSE, SMVITM-Bantakal, Karnataka, India

[4]Student, CSE, SMVITM-Bantakal, Karnataka, India

[5]Asst. Prof, CSE, SMVITM-Bantakal, Karnataka, India

**Abstract**
*A Bio-metrics system is actually a pattern recognition system that utilizes various patterns like iris, retina and biological traits like fingerprint, voice recognition, facial geometry and hand geometry. What makes Bio-metrics really attractive is that the various security codes like passwords and ID cards can be interchanged, stolen or duplicated. To enhance the security and reliability of the system, physiological traits can be used. This paper gives the overview of key bio-metric technologies and basic techniques involved and their drawbacks. Then the paper illustrates the working of ECG and the various opportunities for ECG are also mentioned.*

## 1. INTRODUCTION

Biometrics for authentication: Biometrics traits are the strongest link between a person and his identity as it cannot be easily shared, lost or duplicated. Hence it is more resistant to social engineering attacks. This type of system requires user to be present at the time of authentication. It can deter user from making false claims. Hence it can be incorporated in security applications. Law enforcement agencies in world rely on fingerprints for criminals and forensic identification. Biometric traits deal with access control, checking for multiple access control, international border crossing and secure identification documents. Each biometric trait has its own advantages and weak points. The technique of using biometric methods for identification can be widely applied to forensics, ATM banking, communication security, attendance management systems, and access control. It also plays an important role in enhancing homeland security.Biometric techniques involve 'metrics' or measurements of some kind, rather than depending merely on familiar or hidden methods [8].

Taxonomy of Biometric Techniques given below:
- Appearance - These are the physical descriptions like color of skin or eyes, texture of hair, gender, race, physical markings, height and weight.
- Social behavior -these features correspond to style of speaking, habituated actions and visible handicaps.
- Bio-dynamics - the way, in which a person signs, rhythm of speaking, keystroke dynamics, particularly in relation to login-id and password.
- Natural physiography –These include patterns of fingerprints, hand geometry, retina, iris, DNA, earlobes and many more.
- Imposed physical characteristics like dog-tags, collars, bracelets and anklets, bar-codes, embedded micro-chips [8].

## 2. BACKGROUND

Biometric System: Biometric system is a computer that implements biometric algorithms and makes use of sensing, feature extracting and matching modules. The sensing part captures the traits, feature extraction is done to eliminate unnecessary information and the matching modules match the traits with the references stored in the database. There are two stages in the authentication process. They are the enrollment and verification as shown in Fig 2.1. Enrollment is the process of storing the traits in the database and verification the process of matching the extracted traits with references in stored in database.

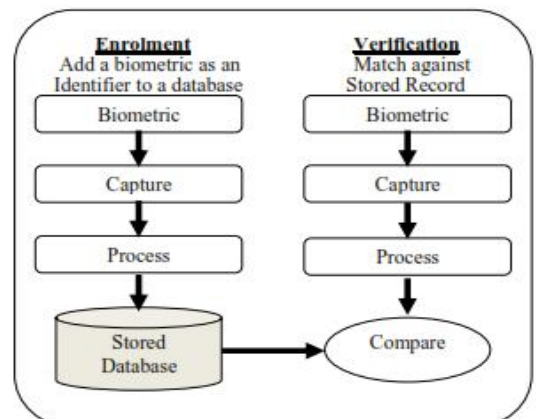

**Fig 2.1:** Biometric System [8]

**Classification of biometric identification systems:**
Biometrics systems is broadly classified into two types

- Physical biometrics: It is based on data derived from direct measurements of parts of the human body. Fingerprints, iris, retina, hand geometry and facial recognition are some of the leading physiological biometrics.
- Behavioral characteristics: Identification of the person is based on his unique behavioral characteristics. These characteristics can be voice, speech, signature, and rhythm of typing or an individual's walking style (gait). Even though the behavioral biometrics is dependent on the actions of an individual, it is also influenced by the physical structure of the human body.

## 3. VARIOUS BIOMETRICS APPROACHES

Following section lists the various Behavioral Biometrics approaches,

- **Keystroke or Typing Recognition:**
  Keystroke dynamics captures the manner or rhythm in which a user types characters on keyboard. The time in which key is pressed is called dwell time and the time interval between "key up" and next "key down" is called as flight time. This timing data is recorded. It is then processed using neural algorithms, which gives us the primary pattern used for future comparisons.
- **Speaker or Voice Authentication:**
  A voice biometric is a numerical representation of the sound, rhythm, and also pattern of an individual's voice. A voice biometric or "voice print," is as unique to an individual as any other biometric methods. Voice authentication is a fairly simple process. To register, a user records the sample of his voice and it is stored in the authenticating system as 'voiceprint'. If the user wishes to access the resource, a sample of their voice is given to the system. A comparison is made

  between the input and voiceprint to validate that the right person is given the access to the resource.

Following section lists the various Physical Biometrics approaches,

- **Fingerprint Identification or Recognition:**
  Fingerprint technology is also known as dactyloscopy. It is a technique of identification where comparison is made between two instances of fingerprints to determine whether they are from same sources. This technique requires comparing several features of fingerprints that are found unique for an individual. The ridges and minutia points are unique for an individual. The ridges can be of three types: loop, whirl and arch. Minutia points are specific points on fingerprints which are critical for identification. These features of fingers are captured as an image with help of scanner and is enhanced and converted to a template. The template is encrypted biometric key or mathematical information. The image of the fingerprint is not stored. The algorithm cannot convert this template back to image. Hence it is difficult to replicate fingerprints. Scientists have discovered that the fingerprints are inherited. It is possible to have a belief that the members of a family share the same fingerprint pattern.
- **Hand or Finger Geometry Recognition:**
  Hand geometry identifies an individual based on the unique features of the hand. The unique features may be length of finger, its thickness, the distance between the finger joints and overall structure of the bone. The system consists of a camera which captures the image of the hand. The necessary features are extracted, processed and stored in the database. These stored templates can be used later for the purpose of verification.
- **Facial Recognition:**
  Face Recognition systems have the ability to identify an individual based on the various features on the face. Human face has several distinguishable features called the nodal points. There are about 80 different nodal points on the face. Some of which are the width of the nose, distance between eyes, length of the jawline, shape of the face based on cheek bones, depth of the eye sockets and many more. The positions of these nodal points are calculated using appropriate algorithms like PCA and LDA. The emerging trend in face recognition is 3D face recognition systems which have better levels of accuracy when compared to the older systems.

## 4. DRAWBACKS OF VARIOUS APPROACHES

Drawbacks of various biometric systems are given below:

- **Fingerprint:**
  Injuries, traumas, wounds or cuts can make the fingerprint reading unidentifiable. Distortions due to grease, dirt or contamination on finger tips have the chances of the identification to be rejected. Today's scanners are still not able to differentiate between real and fake fingerprints. Fake fingerprints can be tricked using gelatinized molds over real finger
- **Facial recognition:**
  Pose variations, ageing are still the limiting factors to identify the person using face recognition. Use of 2D scanners can't handle pose variations and are sensitive to variations in light and shadows. The use of 3D scanners eliminates these limitations but is expensive and has time constraints.
- **Voice recognition:**
  These systems are susceptible to error in the presence of noise and external sounds. The distance of the microphone from the user also affects the accuracy rates. Prerecorded voice can be used for malicious access. These systems take time to adjust to the voice of an individual and require large amount of memory to store voice files.
- **Iris and retinal recognition:**
  The distance of the person from the camera affects the performance of the system. Errors might occur due to reflection caused by spectacles, eyelashes or lenses.

During the scanning process the person is required to remain still. The equipment used is expensive.

- **Hand geometry based recognition:**
  Complications might arise when used with certain populations. There can be a perception of bio- hazard due to spread of potential germs. Possible changes to the shape of hands s can cause failure of the system.
- **Signature**
  These kinds of systems have the limitation factor of inconsistent signature. A person without a constant signature may not be recognized or an individual with muscular illness may face difficulties in proving their identification. The quality of ink and paper may also account for rejections.
- **Keystroke:**
  The major drawback is low accuracy rates due to the varying rhythms of typing. These variations may be due to injury, fatigue, distractions, and mood or due to the side effects of drugs, medications or consumption of alcohol.
- **DNA recognition:**
  As it is a very new technology, may not be very accurate among close relatives and hence less popular in public. Requires time to process and establish identity. Requires expensive equipment for processing and analyzing the samples.

## 5. ECG AS BIOMETRIC TRAIT

ECG based biometric is a recent topic for research. The ECG record is a graphical record of electrical impulses of heart. Electrical activity of the heart is represented by ECG signal. Capturing the ECG signal without the cooperation from the person is a difficult task. Hence cannot be copied easily to provide fake identity. One of the most important strengths of ECG is that it accounts for the vitality of a person. Hence it requires that the person has to be necessarily present at the time of authentication.

All these factors assure better security when compared to other biometric traits which were discussed earlier. Hence ECG can become the most promising biometrics in the near future.

**Description of ECG waveforms:** The ECG is characterized by five different segments of the waveform: the P, Q, R, S and T Waves. [9] The analysis of each segment gives information about different events of the cardiac cycle as shown in fig 5.1.

**P-wave:** It is a small voltage deflection caused due to atrial depolarization.

**QRS-complex:** It is the largest amplitude portion of ECG. It is caused to the currents generated in ventricular depolarization. It is made of three closely related waves: Q wave, R wave and the S wave.

**T-Wave:** It represents ventricular repolarization.

**P-Q interval:** It is the time interval between the beginning of P wave and the beginning of QRS complex.

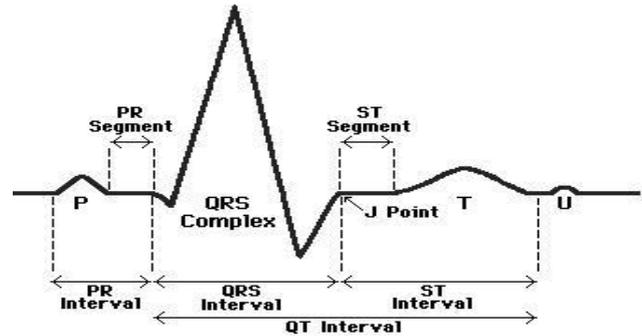

**Fig 5.1** ECG cycle with different segments

**Features that make ECG unique:**
ECG is unique as the morphology and the amplitudes of the cardiac cycle are dependent on the size, shape and position of the heart. The most variable factor of heart is the heartbeat. The normal heart beat of a person is around 60-80 beats/min. This may rise to 200beats/min under pressure or excitement. This variation may reduce the diastole duration and ventricular depolarization. It may also cause the attenuation of the R wave amplitude. But the duration of QRS complex does not vary significantly.

Generally the raw ECG signal is contaminated by noise and hence requires preprocessing to remove the disturbances. Once preprocessing is done, Feature has to be extracted for the purpose of authentication. The features that can be extracted are:

**Angle features:** The angle between PQR, QRS and RST in the ECG signal can be used.

**Interval features:** The time interval between two R peaks can also be used as one of the features. The others are the time difference between a peak and a valley or between two peaks or two valleys.

**Amplitude features:** The difference between the amplitudes of two peaks or two valleys or one peak and a valley can be the features extracted for authentication.

The feature to be extracted has to be selected carefully based on feasibility, consistency and accuracy rates. The angle and interval features are not constant .This because of the fact of the heart beat rate changes from a child to an adult. Hence these features are not suitable. But the amplitude features do not change with age and remain almost constant. Hence the amplitude features may be more suitable for the purpose of authentication.

## 6. APPLICATIONS OF ECG AS A BIOMETRIC TRAIT

We have listed following areas where we can have an authentication process via. ECG.

- In attendance management systems usually ID cards are used for validating the presence of an individual. It is possible for a person to forget his ID at home but very unlikely to forget his heart at home. It is also difficult to make proxy attendance using ECG.
- In e-voting machines, ECG can be used as key factor of single vote. A person can cast single vote. Addition

- of fake votes or multiple votes can easily be identified and rejected.
- In locking systems present in mobile phones, house doors, bank lockers, vehicles etc.
- It can be used in the field of telemedicine to monitor a patient's health over long distances using ECG as identifying factor.
- Used in application involving financial transactions like net banking, ATM systems, online shopping systems etc.
- Used as digital signature to file income tax returns, eProcurement, eTendering, sending and receiving encrypted mails etc.
- May be used in Aadhar cards. Aadhar is 12 digits unique ID issued to every citizen of India. This single source of truth will help in financial inclusions, with deeper penetration of financial institutions and smooth error free distributions of governmental schemes. It card has fingerprints and irises captured. It is likely to fake these biometric factors. But if ECG based identification is used it is difficult to clone the uniqueness of a person.
- Can be used as factor of identification of a person in international border crossing and henceforth avoid the illegal trespassing, smuggling activities and other anti-social activities.
- Gain access control over data files and computer programs which contain personal and crucial information which is of importance to the military, investigations bureaus, revenue, defense and other government organizations. It is very important to maintain secrecy for these agencies.

## 7. CONCLUSION

This paper has evaluated the purpose of using ECG enabled biometric authentication system. Unlike conventional biometrics that is neither secure nor robust enough against falsification, ECG is inherited to an individual is highly secure and impossible to be forged. The most important feature of ECG is its real-time feature of vitality. This ensures that ECG cannot be acquired from a dead person and compulsorily requires the presence of the person at the time of authentication. ECG is the most promising biometric feature due to its uniqueness, universality and acceptability. The characteristic of permanence is also a key factor for it to emerge as an excellent biometric system in the near future